# 有限域LT编码的一种新节点度分布函数

张小弟，樊平毅

（清华大学 电子工程系，北京 100084)

**摘要：** 应用于有限域的Luby Transform (LT)编码为一较新的研究领域。为了探讨有限域LT编码的特点，本文提出一种新的节点度分布函数，思想是希望在相同的译码开销下提高译码成功率，但又不会令编码矩阵失去稀疏性。通过数值仿真的形式验证了新函数的具体效果。结果显示，在有限域LT编码的环境下，随着编码域的增加，本文所提出的新节点度分布函数比Luby提出的节点度分布函数在译码失败率方面有显著的性能增益。因此，本文提出的新节点度分布函数较适合于有限域LT编码中使用。

## Novel Degree Distribution Function for LT codes over Finite Field

CHEONG Siotai, FAN Pingyi

(Department of Electronic Engineering, Tsinghua University, Beijing 100084, China)

**Abstract:** Luby Transform (LT) code over finite field is a recent research topic. In order to find out the properties of LT codes over finite field, a novel degree distribution function is proposed in this paper. The main thinking of our developed distribution function is to improve the decoding success rate with the same overhead, and still to keep the sparse property for the encoding matrix. Numerical simulations are used to show the general performance of our novel function. Various simulation results show that in the environment of LT codes over finite field, our new degree distribution function performs much better than the Luby's degree distribution functions as the field size increasing. In conclusion, our novel degree distribution function is more suitable to be used in LT codes over finite field.

数字喷泉编码[1][2][3]是一类无速率编码，其核心思想是利用一个类似喷泉的编码形式，对任何有限数据量的原始信息文本，可产生无限多个编码数据包，并发送给用户。在用户端，只要正确收集足够数量的数据包，就可以利用特定的解码算法恢复原始信息文本。这样，可使网络资源得到充分利用，同时亦可减少反馈次数。数字喷泉编码的这种无速率编码方式，对于传输链路的特性和信噪比并无特别要求，只要能正确接收到足够的数据包就可进行正确解码。

在数字喷泉编码的概念提出后，具体可实现

收稿日期： 2011-09-19

基金项目： 国家“九七三”重点基础研究项目 (2012CB316100/2012CB316102)

作者简介： 张小弟(1979－)，男（汉），澳门，博士研究生。

通讯联系人： 樊平毅，教授，博士生导师, E-mail: fpy@mail.tsinghua.edu.cn

的编码模式一直没有报道。2002年，Luby给出了一种可行的设计方式，称为Luby转换码，简称LT码[4]，其简洁的编译码设计使之具有良好的性能及低运算复杂度。

LT码除了应用于二元编码外，更可以推广到有限域（或称伽罗华域，Golas Field）上，本文所著重讨论的亦是关于有限域上LT码的性能分析。

及后，于2006年，Shokrollahi在LT码的基础下，提出了Raptor码[5]的概念。它是将LT码与如LDPC编码[6]等分组码作级联编码。由于其独特的设计，巧妙地应用了LT码的特点与LDPC码解码算法的特点，故使编码的复杂度略为增加，但译码的复杂度则保持了与码字长度相匹配的线性增加特征，适合于广播类型的应用。

近期，关于非二元域的数字喷泉编码得到广泛关注。[7]给出了一类设计 G(q)域上数字喷泉编码的模式，[8]讨论了将网络编码与G(q)域上数字喷泉编码的组合设计模式. 有关非二元域数字喷泉编码的研究正在不断扩张。我们知道关于数字喷泉编码的性能，特别是解码性能，与选择的度分布函数有直接关系，而这方面的研究很少有报道。在本文中，我们将探讨有关节点度分布对G(q)域上数字喷泉编码的性能影响，提出了一种新的节点度分布函数，仿真结果表明随着有限域的扩张，其增益越大。

本文的组织结构如下：首先，简述了二元随机线性喷泉码及LT码的特性，以及其编译码方式。及后，介绍应用于有限域中的随机线性喷泉码及LT码的特性。针对应用于有限域的LT编码的特性，本文提出一种新的节点度分布函数，它的特点是在提高译码成功率的前提下，又不会令编码矩阵失去其稀疏性的特点，这有助于利用高斯消元法进行成功译码。之后的部分将会是仿真结果分析，最后对本文进行总结。

## 1 二元随机线性喷泉码

假设一个由K个比特组成的数据文件，分别记为$s_1,s_2,...,s_K$。在第n个时隙内，随机的编码器会随机产生K个二进制编码系数$g_{n,1}$，$g_{n,2}$，...，$g_{n,k}$，相应的编码输出为：

$$u_n=\sum_{k=1}^{K}s_k g_{n,k} \bmod 2 \qquad (1)$$

其中，编码矩阵为$G=\{g_{n,k}\}_{\infty xK}$

假设{ui}为完整接收到的K个部份信息，相应的编码子矩阵$G_K=\{g\}_{KxK}$

利用高斯消元法可得到其逆矩阵，因此:

$$(s_1,s_2,...,s_K)=(u_1,u_2,...,u_K)G_K^{-1} \qquad (2)$$

一般来说，可以通过适当设计的随机矩阵产生方法，使得接收信息数略大于K时，就可以接近1的概率获得一个可逆的子矩阵。额外接收的多余的编码包的数量与码长之比通常被称为译码开销。

对于此类数字喷泉编码，相应的译码开销$\varepsilon$与译码失败率$F$的关系如下：

$$F=1-\prod_{k=1}^{K}\left(1-\frac{1}{2^{K(1+\varepsilon)-k+1}}\right) \qquad (3)$$

## 2 二元LT码

2002年，Luby提出了一种可行的数字喷泉编码方式，称为LT码。Luby总结了之前的设计经验，巧妙地解决了编译码复杂度的问题。

### 2.1 LT码编码方式：

对K个输入符号$x_1,x_2,x_3,...,x_K$，可以产生无限多个输出符号$y_1,y_2,y_3,...,y_K,...$，具体步骤如下：

1）根据预定的节点度的分布，随机选取一个输出符号的度d；

2）利用均匀分布随机选取d个输入符号；

3）相应的输出符号为第2）步选定的d个输入符号的模二和；

4）不断地重覆以上步骤以输出符号，直至接收到接收端反馈“接收完成”的信号为止。

相对于随机线性喷泉码，LT编码实际上只是对每一编码的输出包含多少个原数据文件的数目（即符号的度）作出了一些限制，其中符号的度等于1或2的机率占了50%，目的是配合复杂度较低的译码方式，而相应的编码子矩阵亦为一稀疏矩阵。

### 2.2 LT码译码方式：

置信度传递算法（BP算法）：

1）对所有的校验节点中寻找一个满足如下条件的

校验节点$y_n$:它只与一个输入符号$x_k$相连接。如果条件满足，则令$x_k=y_n$；然后对所有与$x_k$相连的校验节点进行如下赋值运算：$y_i=x_k+y_i$（mod2)，其中$y_i$是任意一个与$x_k$相连的校验节点；最后在图中删除那些与$x_k$相连接的所有的边。如果条件不满足，则运算停止，说明解码失败。
2）重复以上步骤，直至所有$x_k$被确定。

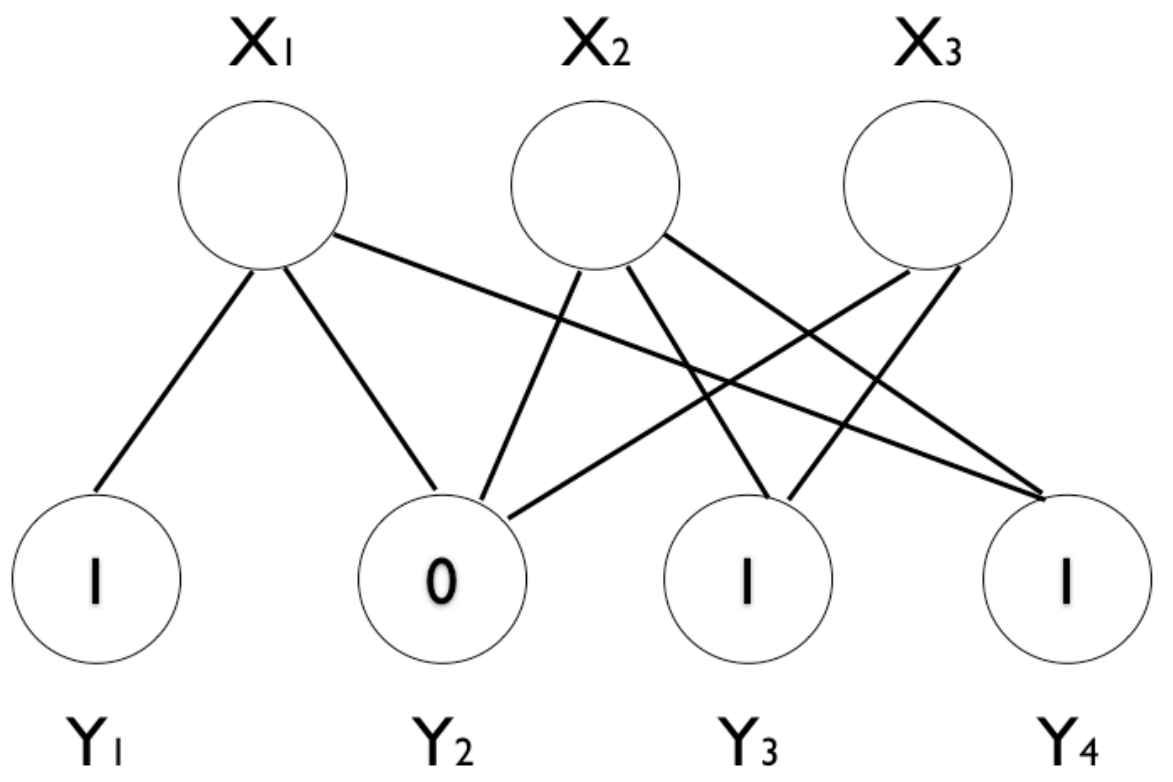

**图1 二元LT编码图，利用BP算法的译码顺序为X1,X2及X3,相应结果为1,0,1。**

置信度传递算法（BP算法）的特点是复杂度较低，一般约为**O**(KlnK),但在没有出现只与一个输入符号连接的校验节点情况下，译码随即失败。与传统的高斯消元法(复杂度约为信息量K的平方[9])相比，失败率会相对较高。在参考文献[9],[10]中亦提出一些改进的高斯消元法去作译码，使之具有较好的译码性能。

从上述步骤可以看出，节点度分布对LT码的译码成败有很大的影响。Luby在文献[4]中提出的松弛型概率分布的修正方法，修正了原来理想的Soliton输出符号节点度分布，使LT码译码更为稳定。其中分别引入两个参数c及$\delta$去调节节点度为1的平均数，从而实现对LT编译码性能的稳定性控制。

Luby给出的鲁帮的 Soliton 度分布函数如下：

$$\tau(d)=\begin{cases}\frac{S}{K}\frac{1}{d} & d=1,2,\ldots,(K/S)-1\\ \frac{S}{K}\ln(S/\delta) & d=K/S\\ 0 & others\end{cases} \quad (4)$$

$S=c\ln(K/\delta)\sqrt{K}$ S即为节点度为1的平均数

除了松弛型概率分布外，亦存在其他种类的节点度分布函数，如Raptor Code[5]中使用的：

$$\Omega_1(x)=0.007969x+0.493570x^2+0.166220x^3+0.072646x^4+0.082558x^5+0.056058x^8+0.037229x^9+0.055590x^{19}+0.025023x^{65}+0.003135x^{66} \quad (5)$$

以上节点度分布表示式中,x的次方数目代表节点度，其系数对应其相应的概率值，为一固定概率的节点度分布函数。在此分布中，节点度为1或2的机会仍占多数，约占50%。

## 3 有限域的随机线性喷泉码

应用于有限域（GF（q),q是质数或质数的正次方）的随机线性喷泉码,其码字由0,1,2,…,q-1组成，而它 的编译码方式大致与二元随机线性喷泉码一致，差别在于原来的运算是在二元域（只有0和1）上，而有限域的运算则是在GF(q)上进行，而所有编码系数均是属于GF(q)。

值得留意的是，在有限域GF(q)中，相应的译码开销$\varepsilon$与译码失败率$F$的关系为：

$$F=1-\prod_{k=1}^{K}\left(1-\frac{1}{q^{K(1+\varepsilon)-k+1}}\right) \quad (6)$$

由此可见，在相同译码开销下，由于q大于2，应用在GF(q)的随机线性喷泉码的译码失败率比二元随机线性喷泉码要低。因此，对于要求高译码成功率的应用，有限域随机线性喷泉码更为适合。而相信应用在GF(q)上的LT码亦会有此特性。

## 4 有限域的LT码

有限域LT码的编码方式与二元LT码大同小异，如下：

### 4.1 有限域LT码编码方式：

在GF(q)中，对K个输入符号$x_1,x_2,x_3,...,x_K$，可以产生无限多个输出符号$y_1,y_2,y_3,...,y_K,...$，具体步骤如下：

1）根据预定的节点度的分布，随机选取一个输出符号的度d；
2）利用均匀分布随机选取d个输入符号；
3）为每一个如此的输入符号，利用均匀分布随机在1,2,3,…,q－1之间选取一相应系数；
4）相应的输出符号为第3）步选定的d个输入符号及其相应系数之积的总和，所有运算均在GF(q)上

进行；

5）不断地重覆以上步骤以输出符号，直至接收到接收端反馈“接收完成”的信号为止。

4.2 有限域LT码译码方式：

可采用BP算法或高斯消元法，针对本文的分析重点，分析将采用传统的高斯消元法，而所有运算均在GF(q)上进行。

4.3 一种新的节点度分布函数：

为了在相同译码开销下，提高LT码译码的成功率，本文提出一种新的节点度分布函数，理念是将固有Raptor码所采用的节点度分布函数与纯随机方式结合，以一半固定，一半纯随机的方法，希望能以其纯随机的特性来增加译码时的成功率，但仍然保持矩阵的稀疏性：

$$\Omega_R(x)=0.007969x+0.493570x^2+0.166220x^R+0.072646x^R+0.082558x^R+0.056058x^R+0.037229x^R+0.055590x^R+0.025023x^R+0.003135x^R \quad (7)$$

其中在（7）式的后八项中，$R$为在[3,4,…,q]中均匀分布中随机选取的节点度随机变数，若各项中R为相同则其概率相加。此式的构筑是以上文提及的概念完成，即将固定的节点度分布函数与纯随机结合。在此新的节点度分布函数中，节点度等于1或2的机率仍然保持约为50%,因此能保持编码矩阵的稀疏性,而余下的50%的机率将由其他节点度所占有，使之具有纯随机性。

在参考文献[11]及[12]中，提及可利用一些概率性方法求出有限域稀疏矩阵的逆矩阵，因此稀疏性有助高斯消元法译码性能提高。

## 5 仿真结果与分析

对于新提出的节点度分布函数的性能，下面我们利用计算机系统仿真探讨其与固有的节点度分布函数性能的差异。根据本文的分析重点，译码方式将采用传统的高斯消元法。在下面的仿真中所采用的鲁帮的Soliton 度分布函数中参数选取如下：c=0.05, $\delta$=0.01和$\delta$=0.001。依据仿真结果,比较目前Raptor码所采用的节点度分布函数和本文新提出的新节点度分布函数, 分析它们对应的译码失败率和译码开销之间的关系。事实上，译码开销的范围是通常选择在0.00至0.05之间。

在下面的仿真中，信息量K为100，q分别为4, 8, 16及32。仿真采用Matlab软件进行，对于每一数据点均进行了10000次独立仿真。为了缩短仿真运行时间，对于所有节点度分布函数，每当因接收到一个新的输出符号（新的译码开销），编码子矩阵产生新的一行时，均会随机替代原有矩阵里的一行，使得编码子矩阵始终保持为方阵，这样可以更好、更快地利用高斯消元法进行译码并对相应的译码失败进行估算。由于对所有节点度的分布函数均进行了上述的操作，因此这种操作并不对它们的性能分析产生影响。

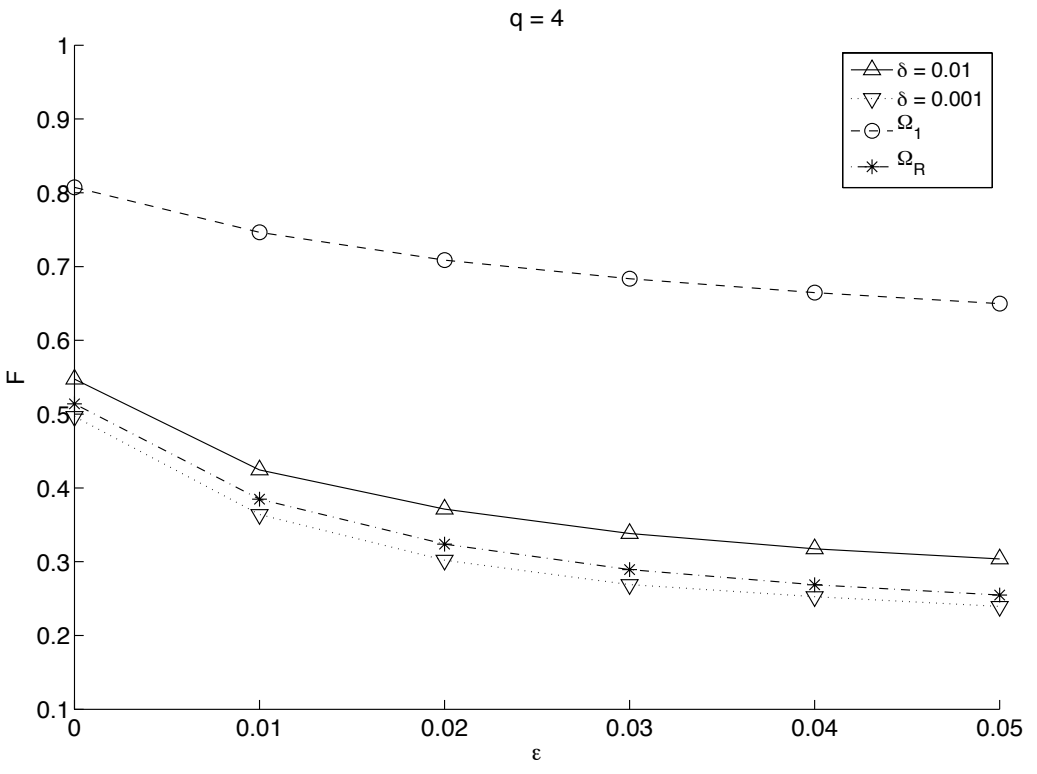

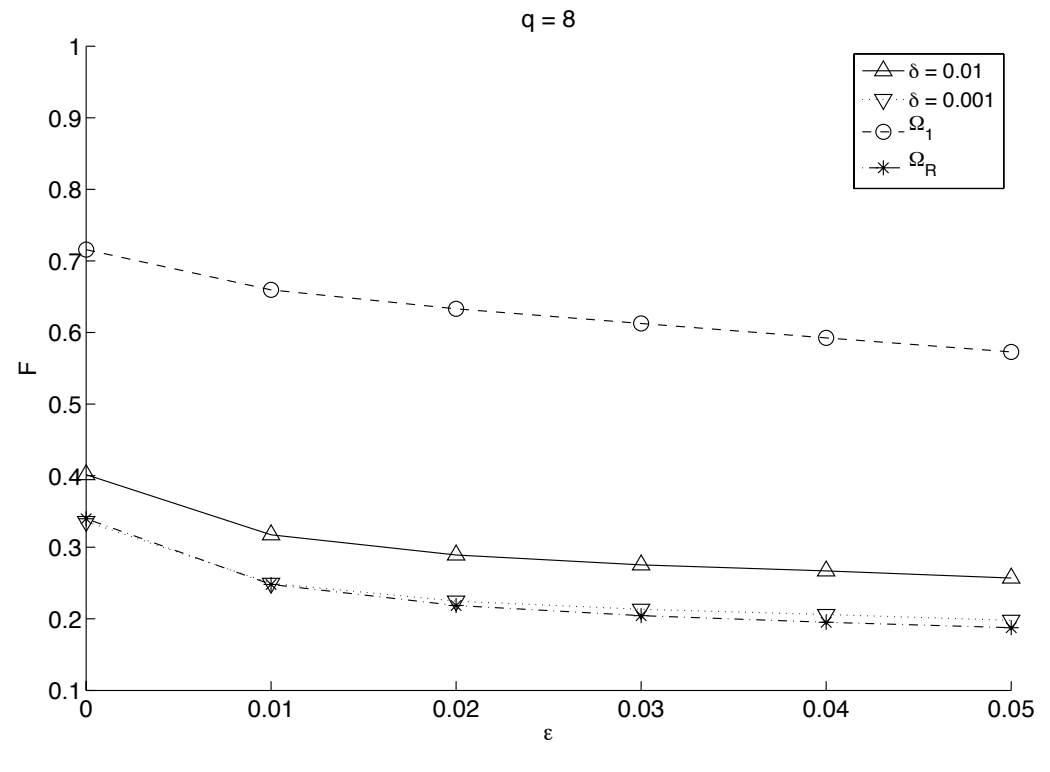

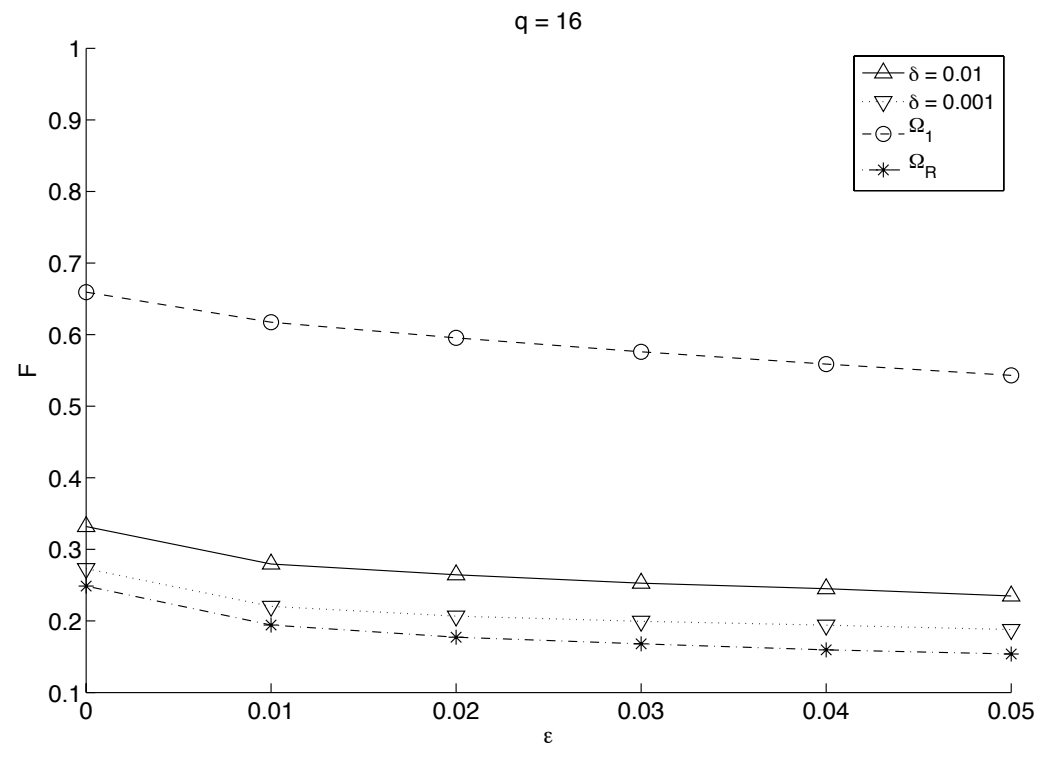

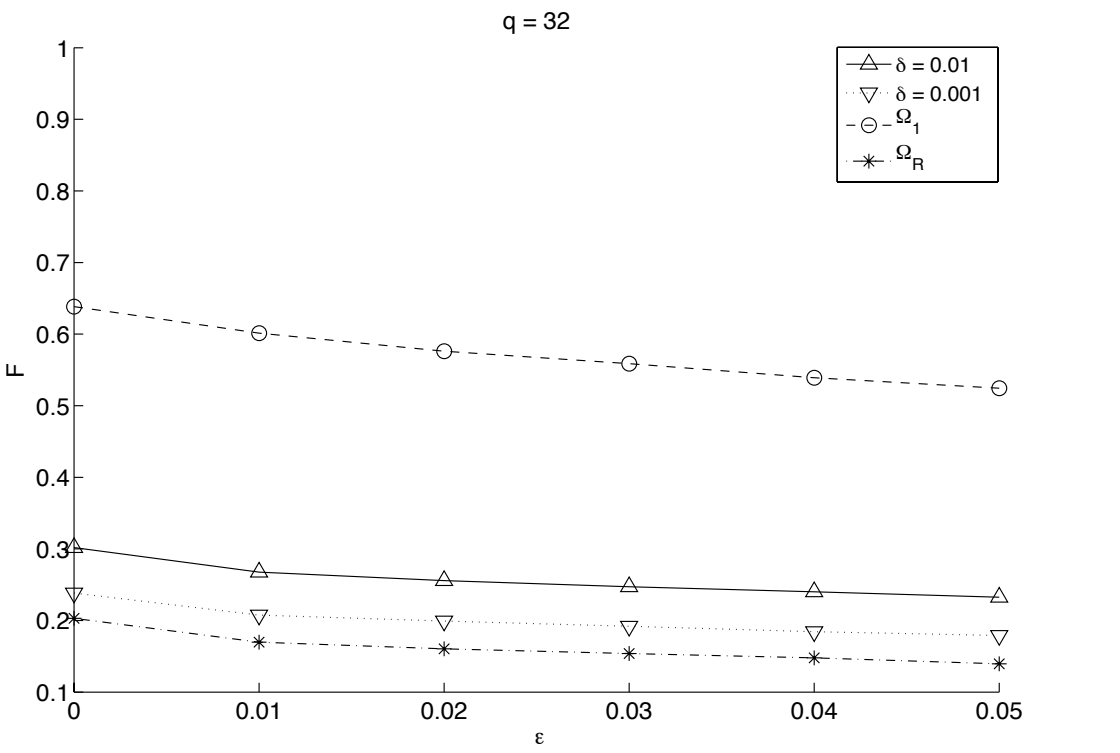

**图2 q=4, 8, 16及32时各节点度分布函数译码开销与译码失败率的关系。**

在上述各图中，横轴代表译码开销 $\varepsilon$ ，取值范围由0.00至0.05，纵轴代表译码失败率$F$。在这里分别以正三角形符号与倒三角形符号代表鲁帮的 Soliton 度分布$\delta$=0.01及$\delta$=0.001两种情况、以圆形代表Raptor所采用的节点度分布以及以星号代表本文新提出的节点度分布。

这些结果表明，随著域值q值的增加，在相同的译码开销的情况下，各种不同节点度分布函数的所对应的译码失败率均下降，系统性能均有提升。而对于相同的q值，随著译码开销的增加，各种节点度分布函数的译码失败率亦随之下降。从以上各图的仿真结果也显示，新提出的节点度分布性能在译码失败率性能上优于固有的Raptor所采用的节点度分布。另一方面， 鲁帮的 Soliton 度分布采用$\delta$=0.001参数时的性能，在固有的节点度分布中是性能最佳的。而在q值大于或等于8时，新提出的节点度分布函数均比原先固有的节点度分布函数在相同的译码开销下有较低的译码失败率。而随著q值的增加，这种性能上的差异越来越大。综以所述，新的节点度分布函数更适合具有较大域值的有限域LT编码中使用。

## 6 结束语

喷泉编码的特点是可以产生无限多个编码数据包，当用户端接收到足够多的数据包时使可以进行译码，适合在各种不稳定的信道上使用。针对非二元有限域LT编码，我们提出了一种新的节点度分布函数算法，并对其进行了仿真分析。仿真结果显示，在较高域值的有限域的LT编码中，本文所提出的创新节点度分布函数在相同的译码开销下具有比目前固有的节点度分布函数有较高的译码成功率。

## 参考文献 （References）

[1] Byers J, Luby M, Mitzenmacher M, et al. A digital fountain approach to reliable distribution of bulk data [J]. *ACM SIGCOMM Computer Communication Review,* 1998, **28**(4):56-67

[2] MacKay D. Fountain codes [J]. IEE *Communication Proceedings,* 2005, **152**(6): 1062-1068

[3] 张小弟，樊平毅. 数字喷泉编码关键技术调研 [J]. 移动通信, 2010, (33a): 17-21.

CHEONG Siotai, FAN Pingyi. Key Technology Research of Digital Fountain Codes [J]. *Mobile Communications*, 2010, (33a): 17-21 (in Chinese)

[4] Luby M. LT codes [C]. *Proceedings of the 43rd Annual IEEE Symposium on Foundations of Computer Science (FOCS02),* Vancouver, Canada: IEEE Press, 2002: 271-280

[5] Shokrollahi A. Raptor codes [J]. *IEEE Transactions on Information Theory,* 2006, **52**(6): 2551-2567

[6] Gallager R. Low-Density Parity-Check Codes [M]. Cambridge: MIT Press, 1963

[7] LI Xiangming, TAO Jiang. Fountain Codes over GF(q) [J]. *Wiley Wireless Communications and Mobile Computing* [OL]. [2011-9-05].http://onlinelibrary.wiley.com/doi/10.1002/wcm.1189/full

[8] Yang S, Yeung W. Coding for a Network Coded Fountain [C]. *IEEE ISIT2011*, Saint-Petersburg, Russia: IEEE Press, 2011: 2583-2587.

[9] Bioglio V, Grangetto M, Gaeta R, et al. On the fly gaussian elimination for LT codes [J]. *IEEE Communications Letters,* 2009, **13**(12): 953-955

[10] Kim S, Ko K, Chung S. Incremental Gaussian elimination decoding of Raptor codes over BEC [J]. *IEEE Communication Letter*, 2008, **12**(4): 307–309.

[11] Wiedemann D. Solving Sparse Linear Equations Over Finite Fields [J]. *IEEE Transactions on Information Theory,* 1986, **IT-32**(1): 54-62

[12] Kaltofen E, Saunders, B. On Wiedemann's Method of Solving Sparse Linear Systems [J]. *Springer Lecture Notes Computer Science*, 1991, (539): 29-38